\documentclass{iaus}
\usepackage{graphicx}

\title[Magneto-seismology of solar atmospheric loops] 
{Magneto-seismology of solar atmospheric \\ loops by means of longitudinal oscillations}

\author[M. Luna-Cardozo, G. Verth \& R. Erd\'{e}lyi]   
{M. Luna-Cardozo$^{1,3}$, G. Verth$^2$ \and R. Erd\'{e}lyi$^3$}

\affiliation{$^1$Instituto de Astronom\'{i}a y F\'{i}sica del Espacio (IAFE), CONICET-UBA, \\ 
CC. 67, Suc. 28, 1428 Buenos Aires, Argentina. Email: {\tt mluna@iafe.uba.ar} \\[\affilskip]
$^2$School of Computing, Engineering and Information Sciences, Northumbria \\
University, Newcastle Upon Tyne, NE1 8ST, UK. Email: {\tt gary.verth@northumbria.ac.uk} \\[\affilskip]
$^3$Solar Physics and Space Plasma Research Centre (SP$^2$RC), University of Sheffield, \\  
Hicks Building, Hounsfield Road, Sheffield S3 7RH, UK. Email: {\tt robertus@sheffield.ac.uk}}

\pubyear{2011}
\volume{286}  
\pagerange{001--003}
\jname{Comparative Magnetic Minima: \\ Characterizing quiet times in the Sun and stars}
\editors{A.C. Editor, B.D. Editor \& C.E. Editor, eds.}
\begin{document}

\maketitle

\begin{abstract}
There is increasingly strong observational evidence that slow magnetoacoustic modes arise in 
the solar atmosphere.  Solar magneto-seismology is a novel tool to derive otherwise directly 
un-measurable properties of the solar atmosphere when magnetohydrodynamic (MHD) wave theory is 
compared to wave observations. Here, MHD wave theory is further developed illustrating how 
information about the magnetic and density structure along coronal loops can be determined by 
measuring the frequencies of the slow MHD oscillations.  The application to observations of 
slow magnetoacoustic waves in coronal loops is discused. 

\end{abstract}

\firstsection 
\section{Introduction}

Damped slow MHD oscillations have been observed in the solar atmosphere 
using high-resolution EUV imager onboard space-borne telescopes (see review by Wang 2011).  
Such oscillations are important because of their potential for the diagnostics of magnetic 
structures by implementation of the method of magneto-seismology, 
through matching the MHD wave theory and wave observations in the solar atmosphere 
to obtain several physical parameters (e.g., magnetic field strength and
density scale height).

The theory of MHD wave propagation in solar magnetic structures initially began
modelling the magnetic structures as
homogenous cylindrical magnetic flux tubes enclosed within a magnetic environment
(Roberts et al. 1984). Later on, more advanced equilibrium models to study slow MHD oscillations 
have also been proposed with, e.g., dissipative effects and gravity (Mendoza-Brice\~{n}o et al. 
2004, Sigalotti et al. 2007), and non-isothermal profiles 
(Erd\'{e}lyi et al. 2008), while the effect of density and magnetic 
stratification had been revisited on transversal coronal loop 
oscillations by Dymova \& Ruderman (2006) and Verth \& Erd\'{e}lyi (2008), respectively.

Here, the governing equation of the longitudinal mode is solved numerically for density  
stratified loops with uniform magnetic field, as well as for expanding 
magnetic flux tubes with uniform density. 
The effect of these stratifications on the frequency ratio of the first overtone to the fundamental
mode is studied.

\section{Governing equation}
The ideal MHD equations are linearized by considering small magnetic and velocity perturbations
about a plasma in static equilibrium [${\vec b} =(b_r, 0, b_z)$ and 
${\vec \upsilon}=(\upsilon_r, 0, \upsilon_z)$, for $r$ and $z$ the radial and longitudinal coordinates, repectively]. 
In the derivation, a uniform kinetic plasma 
pressure is assumed, and the thin flux tube approximation is considered.
The second-order ordinary differential equation governing the longitudinal velocity amplitude is
(see Luna-Cardozo et al. 2012 for a detailed derivation)
\begin{equation}
\frac{d^2 \upsilon_{z}}{d z^2}+ \left(\frac{c_{\mathrm s}^{2}-c_{\mathrm A}^{2}}{c_{\mathrm f}%
^{2}} \right) \frac{1}{B_z} \frac{\partial B_z}{\partial z} \frac{d \upsilon_z}{d z} +
\left[ \frac{\omega^2}{c_{\mathrm T}^2} -\frac{1}{B_z} \frac{\partial^2 B_{z}}{\partial z^2}
-\left(\frac{c_{\mathrm s}^{2}-c_{\mathrm A}^{2}}{c_{\mathrm f}^{2}}
\right) \frac{1}{B_z^2} \left( \frac{\partial B_z}{\partial z}\right)^2 \right] \upsilon_z = 0,
\label{eq1}
\end{equation}
where $c_{\mathrm A}^2= {(B_z^2 /\mu \rho_0)}$, $c_{\mathrm s}^2={(\gamma p_0 / \rho_0)}$, 
$c_{\mathrm f}^{2}=c_{\mathrm s}^2 + c_{\mathrm A}^2$ and 
$c_{\mathrm T}^{2} = (c_{\mathrm s}^{-2} + c_{\mathrm A}^{-2})^{-1}$ are the square of the 
Alfv\'{e}n, sound, fast phase and tube speeds, respectively. In this equation, 
$\omega$ is the angular frequency of the oscillations.

Equation (\ref{eq1}) is numerically solved using the shooting method
based on the Runge-Kutta technique, for density stratified loops 
with uniform magnetic field, as well as for expanding loops with uniform density.
Solar waveguides 
are modelled as axisymmetric cylindrical magnetic tubes with 
tube ends frozen in a dense photospheric plasma at $z=\pm L$.
On average, plasma density and magnetic field strength are expected to decrease with
height above the photosphere (Lin et al. 2004).

\section{Effect of density stratification}

\begin{figure}
   \centering
   \includegraphics[width=6.7cm]{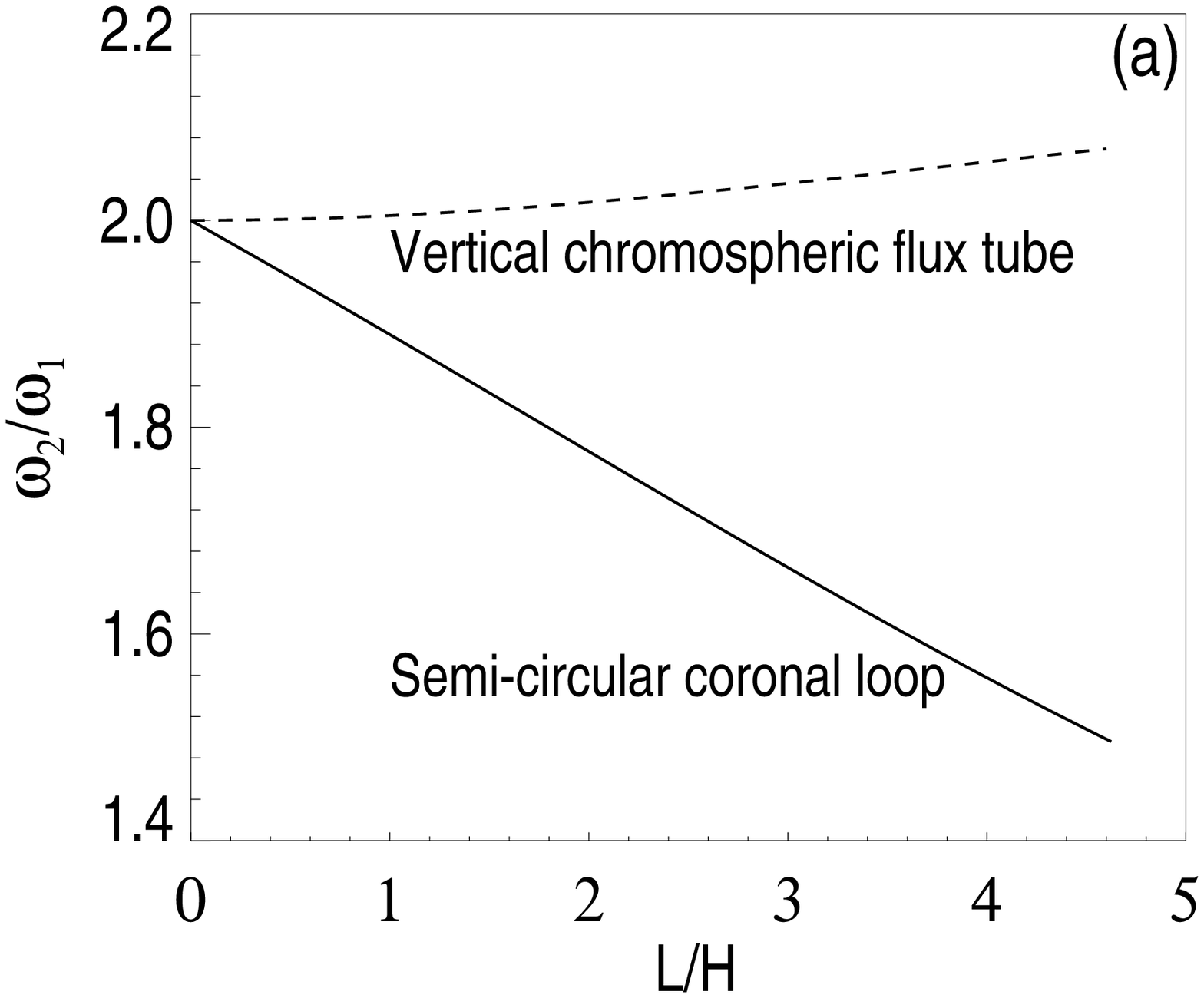}
   \includegraphics[width=6.7cm]{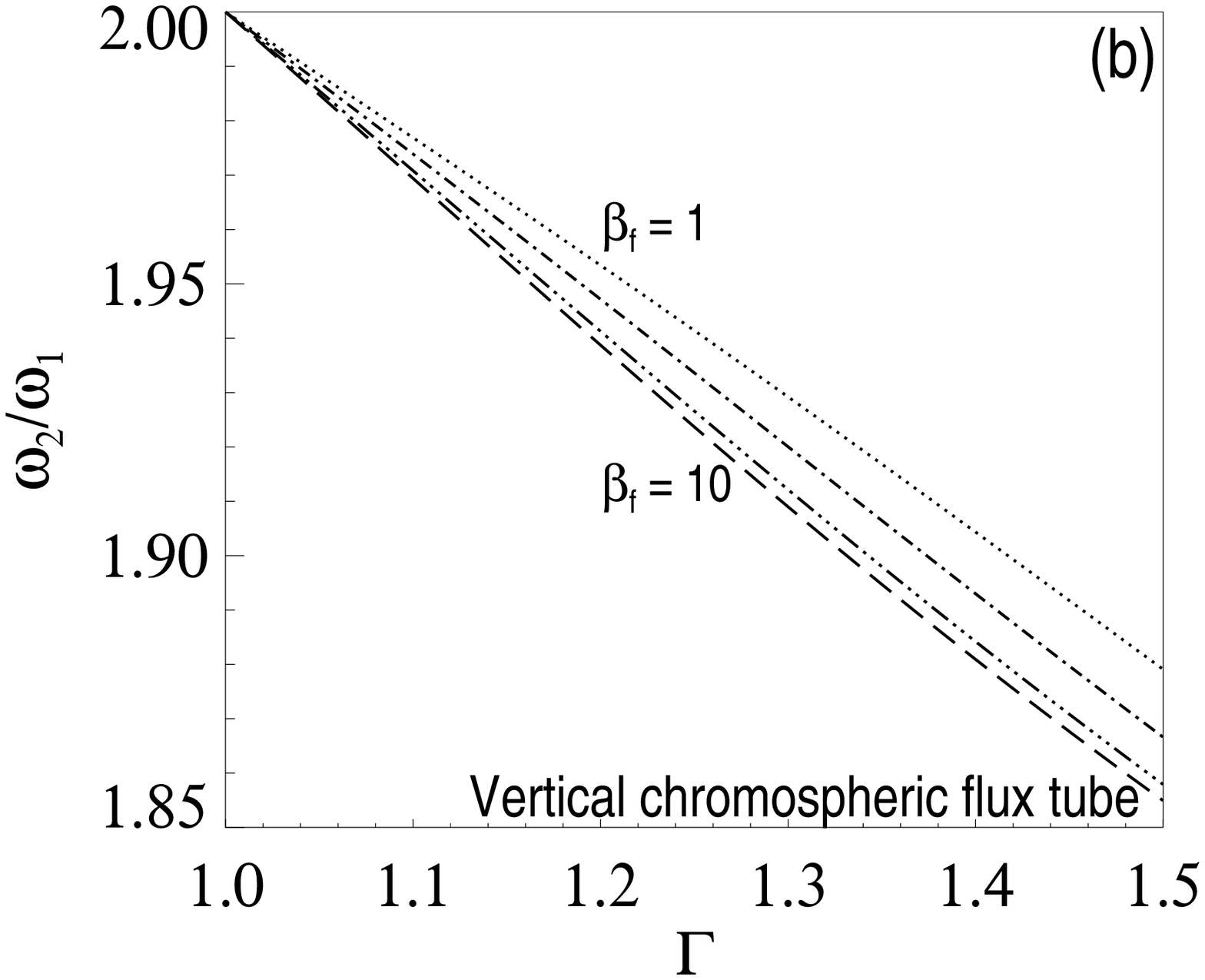}
   \caption{(a) Frequency ratio of the first overtone and fundamental mode 
against $L/H$ for density stratified coronal (solid line) and chromospheric (dashed line) loops. (b) Frequency ratio against the 
expansion parameter $\Gamma$ for different values of $\beta_{\mathrm f}$ in vertical chromospheric flux tubes.
In (b) dotted, dot-dashed, dot-dot-dot-dashed and long-dashed 
lines correspond to $\beta_{\mathrm f} =1$, 2, 5 and 10, respectively.}
           \label{fig1}%
    \end{figure}

The solar coronal loop is modelled by a straight axisymmetric magnetic flux tube with 
tube length of $2L$ and radius of $r_0$. The uniform magnetic field is directed along the tube axis, 
i.e., ${\vec B} = B_z \hat{z}$. In semi-circular coronal loops where the plasma is
close to hydrostatic equilibrium, a reasonable assumption for the density profile is
\begin{equation}
\rho_0 (z) = \rho_{\mathrm f} \exp \left[ -\frac{2L}{\pi H} \cos
\left(\frac{\pi z}{2L}\right)\right], \label{eq2}
\end{equation}
\noindent where $H$ is the density scale height and $\rho_{\mathrm f}$
the density at the footpoint. 
To study a standing wave the boundary condition $\upsilon_{z} (\pm L) = 0$ is applied. 
We solve equation (\ref{eq1}) using the density profile (\ref{eq2}). 
The frequency ratio of the first overtone to the fundamental mode is shown in Figure 
\ref{fig1}(a) as a function of $L/H$ by the solid line, and it is
clearly lower than the cannonical value of two.
A similar result was obtained for the transversal mode by Dymova \& Ruderman 
(2006) and Verth (2007). 

For vertical chromospheric flux tubes the density profile is given by
\begin{equation}
\rho_0 (z) = \rho_{\mathrm f} \exp \left[ -\frac{(z+L)}{H} \right]. \label{eq3}
\end{equation}
Longitudinal oscillations in chromospheric flux tubes are studied solving the 
eigenvalue problem in half of the magnetic bottle, i.e., designating $\upsilon_{z} (-L) = \upsilon_z (0) = 0$ 
as the boundary conditions. 
The ratio of frequencies against $L/H$ for the density
profile (\ref{eq3}) is shown by the dashed line in Figure \ref{fig1}(a).
Now, the frequency ratio is slightly greater than two, indicating that this
parameter depends on the functional form chosen of the equilibrium density.
This suggests that caution must be used when interpreting the frequency ratio 
of chromospheric standing modes. 

\section{Effect of a non-uniform magnetic field}

An expanding flux tube with rotational symmetry about the $z$-axis in 
cylindrical coordinates ($r,\theta,z$) is used to model a magnetic field equilibrium 
decreasing in strength with height above the photosphere.
The magnetic field component $B_z$ at the tube boundary can
be described explicitly as function of $z$ (see Verth \& Erd\'{e}lyi 2008)
\begin{equation}
B_z(z) \approx B_{z,\mathrm{f}} \left\{ 1+\frac{(1-\Gamma^2)}{\Gamma^2}
\frac{\left[\cosh \left(z/L \right)-\cosh(1)\right]}{1-\cosh (1)}
\right\}, \label{eq4}
\end{equation}
where $\Gamma= {r_\mathrm{a}}/{r_\mathrm{f}}$ is the expansion factor, 
and $r_\mathrm{a}$ ($r_\mathrm{f}$) is the apex (footpoint) radius. 
The loop expansion has been estimated for various 
loops, giving mean values of $\Gamma \approx 1.16$ and 1.30 for EUV and soft X-ray 
loops (Watko \& Klimchuk 2000, Klimchuk 2000).

We can compute the
numerical solution of equation (\ref{eq1}) for slow longitudinal oscillations in
coronal and chromospheric loops setting the same boundary conditions as in the previous section, 
and using equation (\ref{eq4}) for $B_z(z)$.  
Figure \ref{fig1}(b) shows the frequency ratio as function of the 
expansion parameter $\Gamma$ for different values of the footpoint 
beta plasma $\beta_{\mathrm f}$ for chromospheric flux tubes. 
It is found that when the
magnetic expansion factor increases the frequency ratio {\it decreases},
and this effect is more significant for chormospheric flux tubes with
higher $\beta_{\mathrm f}$.

\begin{figure}
   \centering
   \includegraphics[width=6.7cm]{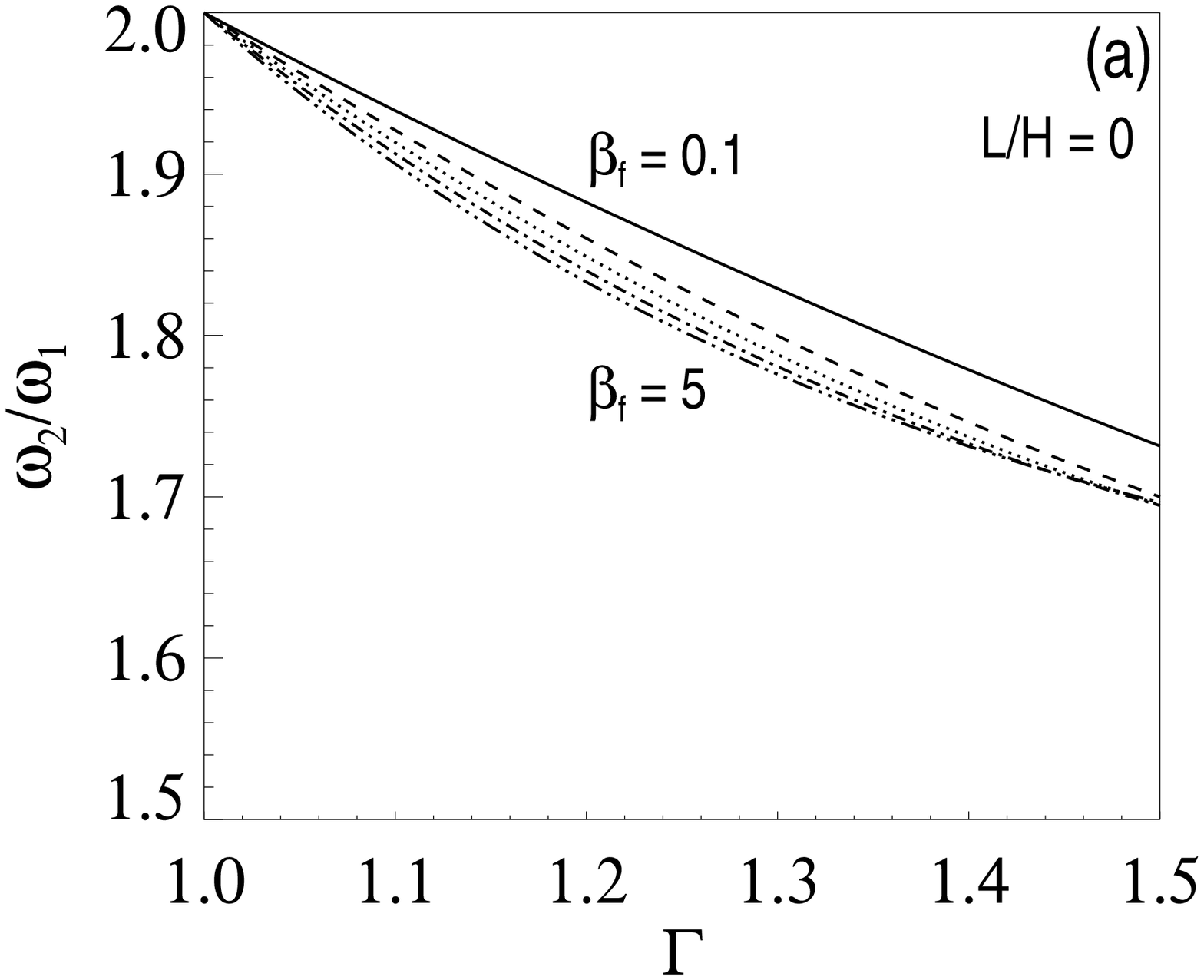}
   \includegraphics[width=6.7cm]{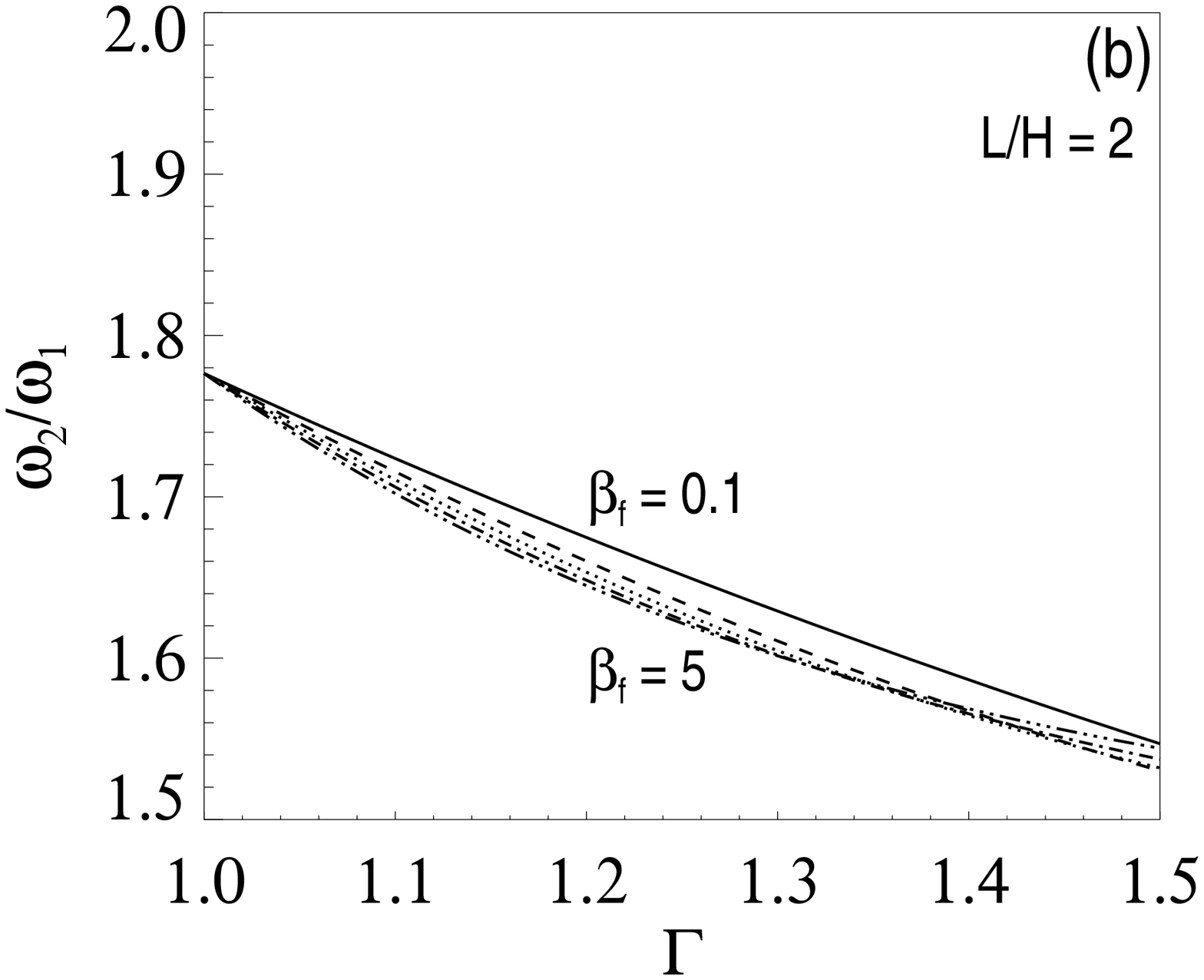}
   \caption{Frequency ratio of coronal loop oscillations against the 
expansion parameter $\Gamma$ for different values of $\beta_{\mathrm f}$.
Solid, dashed, dotted, dot-dashed and dot-dot-dot-dashed 
lines correspond to $\beta_{\mathrm f} =0.1$, 0.5, 1, 2 and 5, respectively.
Coronal loops with uniform density ($L/H=0$) are presented in 
(a) and with typical density stratification ($L/H =2$) in (b).}
           \label{fig2}%
    \end{figure}

Figure \ref{fig2} shows the frequency ratio as function of the 
expansion factor for coronal loops with uniform density in (a) and 
for typical density stratification (i.e., $L/H = 2$) in (b). 
It can be seen how these two effects, density stratification and magnetic expansion, 
contribute to {\it decrease} the value of $\omega_2/\omega_1$.
Additionally, the effect of the expansion is stronger in the corona than in the chromosphere.

\section{Summary and conclusions}

Studying the solutions of the velocity governing equation of the slow standing mode, 
it is found that density stratification and magnetic expansion 
cause the {\it same qualitative effect} on the frequency ratio in coronal loops, giving values of 
$\omega_2 /\omega_1<2$.
For chromospheric flux tubes density stratification and magnetic expansion
cause {\it opposite effects} on the frequency ratio; however, caution
must be taken when studying chromospheric flux tubes since the
ratio $\omega_2 /\omega_1$ depends on the functional form chosen for the
density (see Luna-Cardozo et al. 2012 for an analytical and numerical detailed study about important issues).

These results are consistent with the values of period ratio of $P_1/P_2 = 1.54$ and 1.84 reported by
Srivastava \& Dwivedi (2010) while observing slow acoustic oscillations using {\it Hinode},
in contrast to the theoretical value of $P_1/P_2=2$ for a uniform cylindrical flux tube model.

Our results are important for magneto-seismology, where the density scale height of 
the solar atmosphere can be calculated by using the observed value of the frequency ratio 
$\omega_2 /\omega_1$ 
of longitudinal loop oscillations, to complement both emision 
measure and magnetic field extrapolation studies.
This could provide us with a more complete understanding of the plasma fine structure 
in the solar atmosphere.
These results can be
applied in any stage of the solar cycle, including the solar minimum.
 
\begin{acknowledgements}
M.L.-C. thanks the IAU for the travel grant and is grateful for the financial support from PICT 2007-1790 grant (ANPCyT). 
R.E. acknowledges M. K\'eray for patient encouragement 
and is also grateful to NSF, Hungary (OTKA, Ref. No. K83133) for financial support received.
\end{acknowledgements}

\end{document}